# Internal nanostructure diagnosis with hyperbolic phonon polaritons in hexagonal boron nitride

*Siyuan Dai [†], Mykhailo Tymchenko [†], Zai-quan Xu [‡], Toan T. Tran [‡], Yafang Yang [§], Qiong Ma [§], Kenji Watanabe [∥], Takashi Taniguchi [∥], Pablo Jarillo-Herrero [§], Igor Aharonovich [‡], Dimitri N Basov [Δ], Tiger H Tao [▫], Andrea Alù [†±×▽\*].*

[†] Department of Electrical & Computer Engineering, The University of Texas at Austin, Texas 78712, USA

[‡] Institute of Biomedical Materials and Devices, Faculty of Science, University of Technology Sydney, Ultimo, New South Wales 2007, Australia

[§] *Department of Physics, Massachusetts Institute of Technology, Cambridge, Massachusetts 02215, USA*

[∥] *National Institute for Materials Science, Namiki 1-1, Tsukuba, Ibaraki 305-0044, Japan*

[Δ] Department of Physics, Columbia University, New York, New York 10027, USA

[▫] Department of Mechanical Engineering, The University of Texas at Austin, Texas 78712, USA

[±] Photonics Initiative, Advanced Science Research Center, City University of New York, New York 10031, USA

▫ Physics Program, Graduate Center, City University of New York, New York 10016, USA

▫ Department of Electrical Engineering, City College of The City University of New York, New York 10031, USA

* Correspondence to: aalu@gc.cuny.edu

**ABSTRACT:**

**Imaging materials and inner structures with resolution below the diffraction limit has become of fundamental importance in recent years for a wide variety of applications. In this work, we report sub-diffractive internal structure diagnosis of hexagonal boron nitride by exciting and imaging hyperbolic phonon polaritons. Based on their unique propagation properties, we are able to accurately locate defects in the crystal interior with nanometer resolution. The precise location, size and geometry of the concealed defects is reconstructed by analyzing the polariton wavelength, reflection coefficient and their dispersion. We have also studied the evolution of polariton reflection, transmission and scattering as a function of defect size and photon frequency. The nondestructive high-precision polaritonic structure diagnosis technique introduced here can be also applied to other hyperbolic or waveguide systems, and may be deployed in the next-generation bio-medical imaging, sensing and fine structure analysis.**

Polaritons are hybrid light-matter modes coupling free-space photons and collective resonances supported by a broad range of materials, including metals, semiconductors and insulators[1]. Van der Waals (vdW) crystals – layered materials where atomic planes are bonded together by weak vdW forces – are emerging as a new platform for polaritonic phenomena[2, 3]. Polaritons in vdW crystals and their heterostructures reveal exceptional optical confinement[4-14], mechanical and/or electrostatic tunability[5, 6, 15] and long propagation lengths[11]. These unique properties enable applications including bio-chemical sensing[16-18], lasing[19, 20], wavefront shaping[21] and sub-diffractive focusing[12, 22, 23]. One representative vdW system for nano-polaritonics is hexagonal boron nitride (hBN)[9, 24-33], where anisotropic phonon resonances support Type I ($\varepsilon_z < 0$, $\varepsilon_{xy} > 0$) and Type II ($\varepsilon_{xy} < 0$, $\varepsilon_z > 0$) hyperbolic responses inside the lower and upper Reststrahlen band, respectively. The hybrid photon-lattice modes in hBN therefore propagate as guided waves and are referred to as hyperbolic phonon polaritons (HPPs). In this work, we harness HPPs in hBN to diagnose concealed structure in a fashion similar to X-ray tomography[34], with a resolution down to few nanometers at infrared (IR) frequencies. We investigate the interaction between propagating polaritons and the internal structure of polaritonic media by extracting reflection, transmission and scattering coefficients of HPPs therein. Analysis of the polariton wavelength, dispersion and reflection at different regions reveals the precise location, geometry and size of concealed defects, and therefore enables the trustworthy reconstruction of the internal structure[35, 36] of nanoobjects under investigation.

Our polaritonic defect diagnosis is based on IR nano-imaging, using scattering-type scanning near-field optical microscopy (s-SNOM, Figure 1a). By illuminating a metalized atomic force microscope (AFM) tip with an infrared (IR) laser, we generate strong electromagnetic near fields underneath the apex, which act as a localized optical antenna[37]. These fields provide a wide range of momenta ($q$), and therefore facilitate energy transfer and momentum bridging from free-space photons to propagating HPPs. As established in previous works, the s-SNOM tip acts both as launcher[26] and detector of HPPs in hBN, when samples are scanned underneath. In our experiment, the hBN edge (Figures 1b-d, white dashed lines) reflects propagating HPPs launched by the tip, and the standing waves induced by these interactions are recorded as oscillation fringes in the s-SNOM amplitude $s(\omega)$ (Figure 1b) image. These fringes exhibit the strongest oscillation close to the hBN edge, followed by weakly damped ones away from the edge. The period of polariton fringes equates one half of the in-plane HPP wavelength $\lambda_p/2$.

In addition to these expected features, we observe another series of HPP fringes in the hBN interior: these fringes appear on both sides along the white dotted trace (Figure 1b). Although these polariton fringes show weaker oscillations compared to those close to the edge, they exhibit the same overall pattern: the strongest oscillation followed by further damped waves away from the white dotted traces. While these additional polariton fringes are evident in our experiments (Figures 1b-d), they are not associated with topographic features in the AFM image simultaneously recorded from the same sample (Figure 1e). Interestingly, HPPs exhibit distinct wavelengths at various locations in the same hBN slab (Figure 1b): polariton fringes along the hBN edge (white dashed line) are bent towards the edge from Region 1 to Region 2. Line traces (blue, black and red dashed lines in Figure 1b), taken as cuts from the s-SNOM image, present polariton fringes as oscillation peaks in Figure 2a: red fringes (hBN 2) are closer to the crystal edge ($L = 0$) compared with the blue ones (hBN 1). The associated polariton wavelength can be measured as the distance between fringe peaks (arrows): the wavelength $\lambda_p$ of HPPs in Region 1 (blue and black arrows) is longer than the one in Region 2 (red arrow). This wavelength difference can also be observed at other IR frequencies (Figures 1c-d), and it persists over a broad frequency ($\omega$) – momentum ($q$, $q$

= $2\pi / \lambda_P$) space (Figure 2b) within the Type II hyperbolic region. Data extracted from the hBN region 1 (blue dots in Figure 2b) agree well with our simulations (green curve, see Supplementary Section 1 for details) using the hBN thickness (157 nm) obtained from the AFM topographic image (Figure 1e). Since the wavelength of HPPs scales linearly with the thickness of hBN,[9, 25] and there is no topographic difference between Region 1 and Region 2 (Figure 1e), we deduce that an air gap was formed within the hBN Region 2 (Figure 1a). This air gap can occur if numerous constituent layers of hBN crystal have been removed – for instance due to mechanical damage during exfoliation (Methods). The edge of the air gap (white dotted line) therefore acts as a polariton reflector in the hBN interior, and thus produces the interference fringes detected in the polariton images.

We now show that the vertical position ($z$) and thickness ($d$) of the interior void can be precisely extracted by analyzing the HPP wavelength (Figure 3) in the polariton images. Note that previous works have demonstrated resolving subsurface objects[38, 39] using spatially extended evanescent near-fields generated by the s-SNOM tip, despite a reduced imaging resolution away from the surface. Permittivity and topographic information have also been retrieved on thin isotropic materials by analyzing s-SNOM data at multiple harmonics with detailed tip-sample interaction model[40-42]. However, the retrieved information remains valid within a limited depth below the sample surface (< 50nm) due to the evanescent nature of the optical fields, unless it is coupled to volume confined propagating modes. Here, highly confined polaritons in hBN slabs propagate as guided-waves[9, 24] in the hBN Region 2 and is thus sensitive to the internal structure of the slab: the polariton wavelength $\lambda_P$ is affected by the thickness $d$ and the vertical position $z$ of the air gap. Therefore, the analysis of HPPs and its wavelength $\lambda_P$ can be used to extract $z$ and $d$ without limitation very close to the sample surface ($z = 0$). At a representative IR frequency ($\omega$ = 1515 cm⁻ [1]), our simulation (Supplementary Section 1) shows that $\lambda_p$ decreases with increasing $d$, while, at a fixed $d$, $\lambda_p$ first decreases and then increases with the increasing vertical position $z$ (see color curves in Figures 3a-b). Simulation curves that intersect the black dotted line (Figure 3a) can fit our experimental results ($\lambda_p$ = 1215 nm, black arrow). Based on the fitting results at $\omega$ = 1515 cm$^{-1}$, we can estimate the range of the void thickness $d$ = 15 ~ 27 nm. Following this analysis at multiple IR frequencies (for example, at another IR frequency $\omega$ = 1504 cm$^{-1}$ shown in Fig. 3b), we can narrow the range of the void position and thickness by overlapping all the estimated ones. We finally conclude that the air gap in Region 2 locates at $z$ = 111 ~ 113 nm with a thickness $d$ = 18 ~ 23 nm.

Having obtained the tomographic information of the defective hBN slab with our analysis, we now examine the propagation properties of HPPs, including polariton reflection, transmission and scattering at the internal void inside hBN. The imaging of polariton fringes in hBN (Figure 1) is based on the reflection of polaritons at the edge or at the internal defect. As the polariton reflection is directly related to the fringe oscillation amplitude[43, 44], we can witness a difference in reflection amplitude of polaritons at the hBN edge and at the inner defect in Figures 1b-d. We systematically extracted the reflection amplitudes of polaritons (Figure 4a) at the inner defect by analyzing the HPP oscillation amplitude from the hBN Region 1 (blue dots) and from the hBN Region 2 (red squares). The reflection amplitude can be estimated as $R(\omega)_{1,2}$ = ($S(\omega)_{m\_defect\ 1,2}$ − $S(\omega)_{background\ 1,2}$) / ($S(\omega)_{m\_edge\ 1,2}$ − $S(\omega)_{background\ 1,2}$) in hBN region 1 and 2. $S(\omega)_{m\_defect}$ and $S(\omega)_{m\_edge}$ are the s-SNOM amplitude at the strongest polariton fringe close to the internal defect and the crystal edge, respectively and $S(\omega)_{background}$ is the background s-SNOM amplitude of the sample. Note that $S(\omega)_{background\ 1}$ and $S(\omega)_{background\ 2}$ are different, we therefore normalize the reflection amplitudes at the internal defect to those at the crystal edge, provided that all polariton fringes are registered equally (Figure 1a), and the latter reflection amplitudes are close to unity. To support our experimental results, we performed finite element simulations of HPPs in a defective hBN slab (Figure 4c) using COMSOL Multiphysics (Supplementary Section 2). The simulated polariton reflection with various air gap thicknesses are plotted in Figure 4a as solid curves. Our experimental data are shown in the same panel and fit well with the simulations assuming $d$ = 15 ~ 25 nm – confirming the robustness of our diagnosis technique to extract topographic information of the polaritonic structure (Figure 3). In addition to polariton reflection $R$, the extraction of transmission $T$ and scattering $S$ at the internal defect is possible, but it requires deposition of polariton launchers on hBN, which may be explored in future experiments.

Based on our numerical simulations, we show the evolution of polariton reflection ($R$), transmission ($T$) and scattering ($S$) into polaritons belonging to other hyperbolic branches with frequency $\omega$ at the inner defect (Figure 4b). This analysis uncovers two important results: first, polaritons are more likely to be reflected from thicker defects due to the larger reflection cross-section; second, as the frequency $\omega$ increases, polaritons are more likely to be reflected because the defect thickness becomes more comparable to the polariton wavelength ($\lambda_p$ / $d$ decreases). A direct experimental evidence of this second point is that the internal defect is less evident at $\omega$ = 1525 cm$^{-1}$ (Figure 1d) compared to the images at higher frequencies (Figures 1b-c): the fringe amplitude at the internal defect in Figure 1d is small, as HPPs are only weakly reflected at $\omega$ = 1525 cm$^{-1}$. Finally, we note that the reflection phase[44] of polaritons at the hBN edge and at the inner defect is different, as revealed in the lateral position of the fringe peaks (blue and black dashed arrows) in Figure 2a: while they have identical wavelength $\lambda_p$, the HPP fringes are closer to the internal defect than to the hBN edge.

The experimental and theoretical results presented in Figures 1 to 4 describe our polaritonic structure diagnosis scheme: detection and imaging of concealed internal nanostructures utilizing hyperbolic polaritons. The core principle of this technique is based on hyperbolic polariton interferometry originating from reflection, transmission and scattering of polaritons at the inner structure/defect. This technique is inherently nondestructive, and it can provide the precise position, size and geometry of the inner structure or buried objects with a resolution down to few nanometers. The idea of harnessing hyperbolic phonon polaritons in hBN for structure diagnosis can be extended to optical waveguides[45], other nano-polaritonic and hyperbolic systems, including black phosphorus[46], topological insulators[47], transition metal dichalcogenides and metal-insulator waveguides[48-50]. We envision future efforts towards exploring the fingerprint of concealed nano-objects/defects based on their interaction with the deeply subwavelength features of polaritons affected by their specific material properties, such as polarizability and conductivity. The nondestructive technique of polaritonic structure diagnosis introduced here, with its nanoscale sensitivity, may benefit the next-generation of bio-medical imaging, sensing and fine structure analysis.

**Methods**

**Experimental setup**

The polaritonic structure diagnosis described in the main text were performed using a scattering-type scanning near-field optical microscope (s-SNOM). The s-SNOM used in our experiment is a commercial system (www.neaspec.com) based on a tapping-mode atomic force microscope (AFM). We use a commercial AFM tip (tip radius ~ 10 nm) with a $PtIr_5$ coating to launch and detect propagating polaritons. The AFM tip is illuminated by monochromatic quantum cascade lasers (QCLs) (www.daylightsolutions.com) that can cover a frequency range of 900 – 2300 $cm^{-1}$ in the mid-infrared. Our s-SNOM signal was recorded by a pseudo-heterodyne interferometric detection module with an AFM tapping frequency 280 kHz and tapping amplitude around 70 nm. In order to subtract background signal, we demodulate the s-SNOM output signal at the 3$^{rd}$ harmonics of the tapping frequency.

**Sample fabrication**

Hexagonal boron nitride (h-BN) crystals were mechanically exfoliated from bulk samples and deposited onto Si wafers capped with 285 nm thick $SiO_2$.

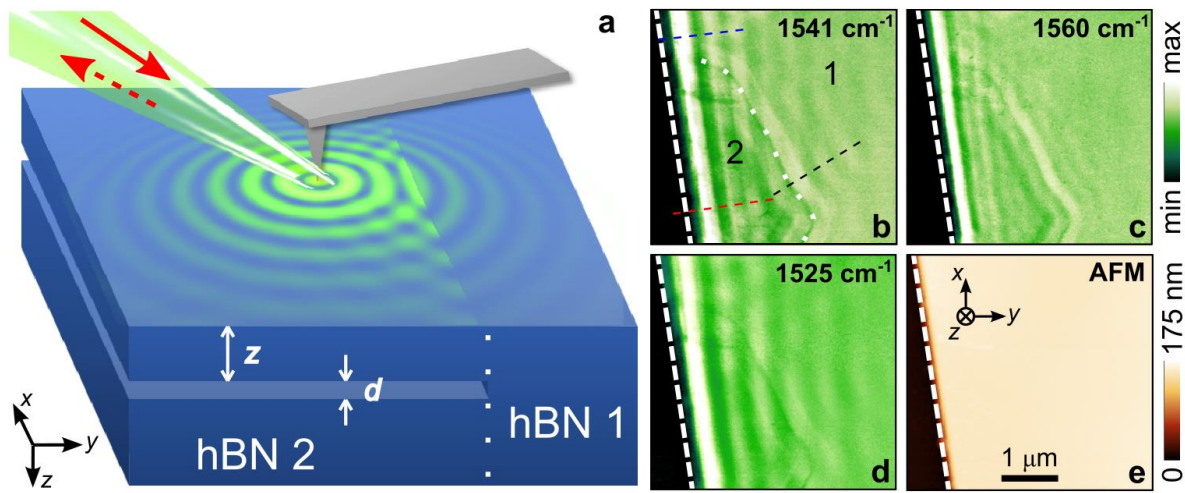

**Figure 1.** Schematic and images of the polaritonic structure diagnosis. **a**, Experiment setup. The AFM tip is illuminated (solid red arrow) by and an infrared (IR) beam from QCL. We collect the back-scattered IR signal. The inner defect (air gap) acts as reflector for the propagating polaritons. **b-d**, s-SNOM images of the hBN slab revealing concealed inner defects at ω = 1541 cm$^{-1}$ (b) 1560 cm$^{-1}$ (c) and 1525 cm$^{-1}$ (d). **e**, Simultaneously recorded AFM image, showing no topographic features. Scale bar: 1 μm.

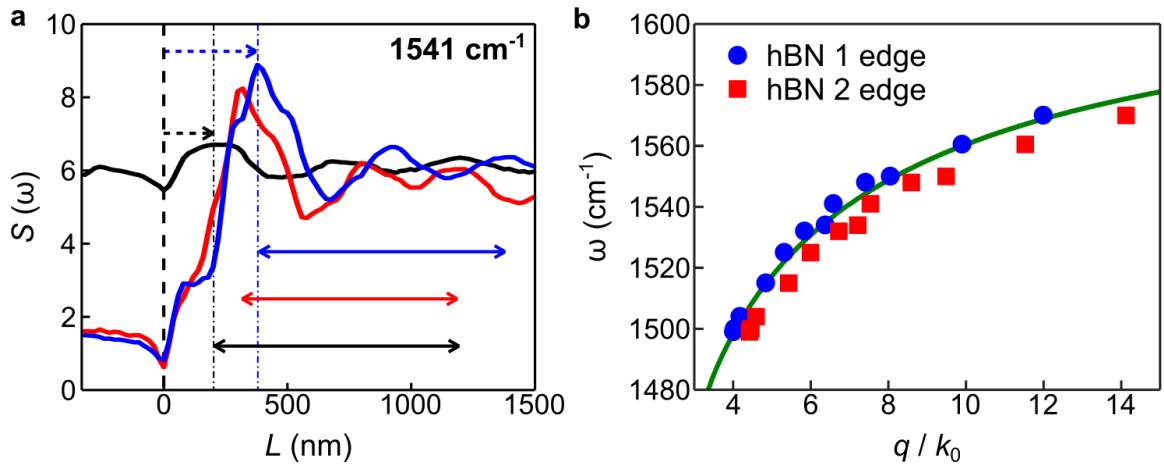

**Figure 2.** Polariton line traces and dispersion. **a**, s-SNOM line traces taken along the cuts in Fig. 1b, the corresponding polariton wavelengths are indicated with double arrows. Single arrows indicate the position of the first fringe peak. IR frequency ω = 1541 cm$^{-1}$. **b**, Frequency – momentum dispersion extracted from fringes along the crystal edge in hBN region 1 and region 2 (Fig. 1b). The experimental data are indicated with blue dots and red squares whereas the simulation results (in hBN region 1) are plotted with the solid green line. $q = 2\pi / \lambda_p$, $k_0 = 2\pi / \lambda_0$, where $\lambda_p$ and $\lambda_0$ are the wavelength of polariton and IR illumination.

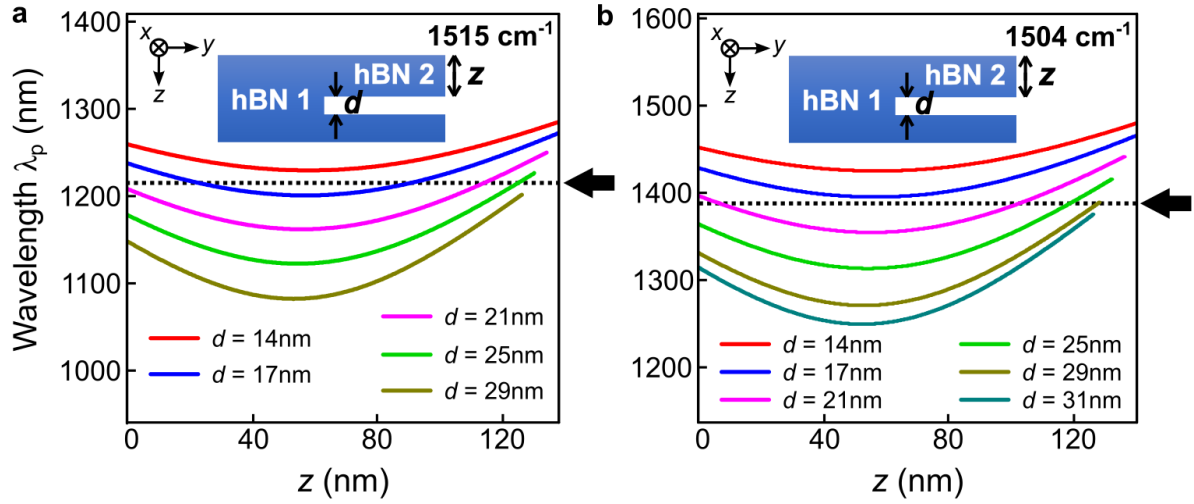

**Figure 3.** The dependence of polariton wavelength $\lambda_p$ (hBN region 2) on air gap position $z$ and thickness $d$ at $\omega = 1515$ cm$^{-1}$ (**a**) and 1504 cm$^{-1}$ (**b**). Black arrow indicates our experimental result $\lambda_p = 1215$ nm at $\omega = 1515$ cm$^{-1}$ (**a**) and $\lambda_p = 1388$ nm at $\omega = 1504$ cm$^{-1}$ (**b**). Color traces are the simulation results with different gap thickness ($d$ = 14, 17, 21, 25, 29 and 31 nm).

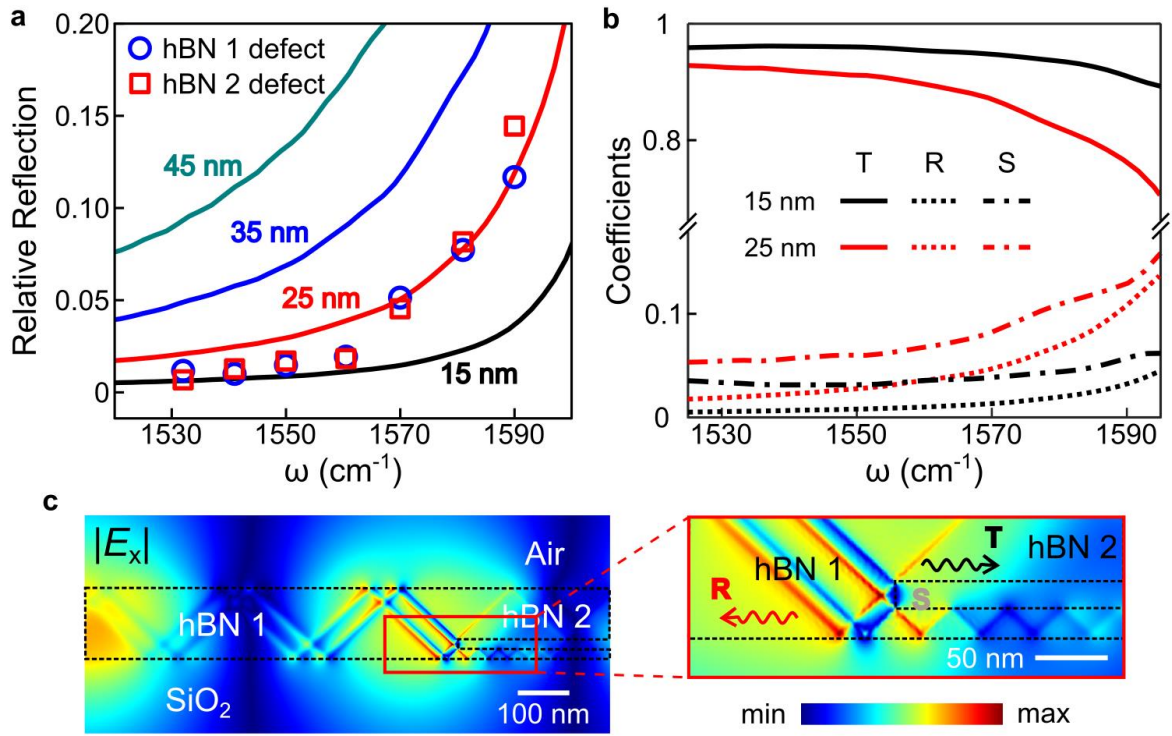

**Figure 4.** Polariton reflection, transmission and scattering at the interior defect. **a**, Evolution of normalized polariton reflection at the inner defect with the IR frequency. Experimental data are plotted with hollow blue dots (in hBN region 1) and red squares (in hBN region 2) whereas simulation results for different thicknesses ($d$ = 15, 25, 35 and 45 nm) of the gap are plotted with color curves. All the reflection coefficients are normalized to those at the crystal edge. **b**, The evolution of polariton reflection $R$, transmission $T$ and scattering $S$ at the inner defect with IR frequency for air gap thickness $d$ = 15 nm (black) and 25 nm (red). **c**, Left, the $|E_x|$ simulation of polariton propagating at the inner defect in hBN. Right, zoom-in simulation marked on the left panel.

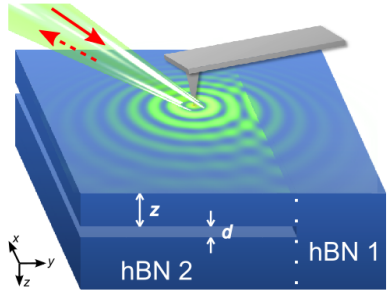

**Table of Contents.**

AUTHOR INFORMATION

**Corresponding Author**

* Email: aalu@gc.cuny.edu (A.A.)

ACKNOWLEDGMENT

S.D supported by the AFOSR MURI grant number FA9550-17-1-0002 and the Welch Foundation with grant No. F-1802. Work at Columbia University on optical phenomena in vdW materials is supported by the Gordon and Betty Moore Foundation's EPiQS Initiative through Grant GBMF4533 and AFOSR FA9550-15-1-0478. P.J-H acknowledges support from AFOSR grant number FA9550-16-1-0382.

ABBREVIATIONS

s-SNOM, scattering-type scanning near-field optical microscopy; hBN, hexagonal boron nitride; vdW, van der Waals; HPP, hyperbolic phonon polariton; IR, infrared; AFM, atomic force microscope.

SUPPORTING INFORMATION

Simulation of hyperbolic phonon polaritons in hexagonal boron nitride with/without the internal defect; Numerical modeling of polaritons and reflection, transmission and scattering.